\documentclass[twocolumn,showpacs,prl,superscriptaddress,]{revtex4}
\usepackage[british,UKenglish,USenglish,english,american]{babel}
\usepackage{amsmath}
\usepackage{amssymb}
\usepackage{latexsym}
\usepackage{enumerate}
\usepackage{amssymb}
\usepackage{graphicx}
\usepackage{slashed} 
\usepackage{xcolor}
\usepackage{array,tabularx,booktabs}
\usepackage{xspace}
\usepackage{multirow}

\begin{document}

\title{Tomographic image of the proton}


\author{R. Dupr\'e$^{1}$,  M. Guidal$^1$ \& M. Vanderhaeghen$^2$\\
\textit{$^1$Institut de Physique Nucl\'eaire d'Orsay, CNRS-IN2P3,\\ 
Universit\'e Paris-Sud, Universit\'e Paris-Saclay, 91406 Orsay, France.}\\
\textit{$^2$Institut f\"ur Kernphysik and PRISMA Cluster of Excellence,
Johannes Gutenberg-Universit\"at, Mainz, Germany.}
}

\begin{abstract}
We determine, based on the latest experimental Deep Virtual Compton Scattering experimental data, 
the dependence of the spatial size of the proton on the quark's longitudinal momentum.
This results in a three-dimensional momentum-space image and tomography of the proton.
\end{abstract}

\maketitle

More than 50 years after the discovery of the partonic substructure of the proton, 
the precise way in which the quarks and gluons compose the nucleon and build up its global properties,
i.e. its mass, momentum, charge, or spin distributions is still not well-known and understood.
The past two decades have seen an important progress both theoretically and experimentally in exploring 
proton structure through the $ep\to ep\gamma$ process, or Deeply Virtual Compton Scattering (DVCS). 
The angular and energy distributions of the scattered electron and radiated photon
reflect both the momentum and space distributions of the quarks within the proton. In the present work, we 
perform a global analysis of recent DVCS data and extract the transverse extension of the proton for different 
longitudinal quark momentum slices. 

\begin{figure}[h]
\begin{center} 
\includegraphics[width=6cm]{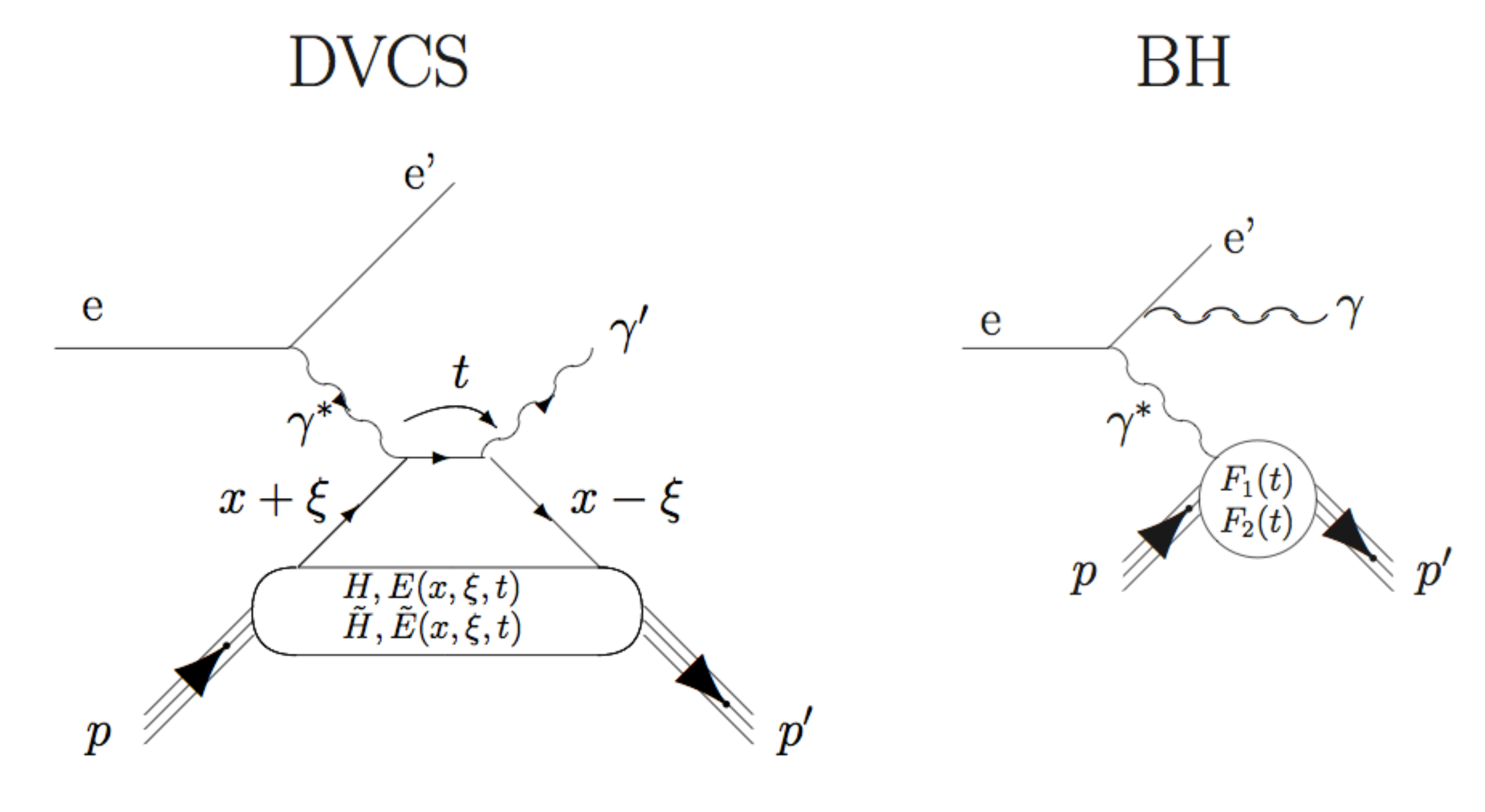}
\vskip -.25cm
\caption{Left: the DVCS process (there is also a crossed diagram where the
final state photon is emitted from the initial quark). Right: the BH process
(there is also the process with final state photon emitted from the
initial electron). The various variables and quantities are defined in the text.}
\label{fig:dvcsbh}
\end{center}
\end{figure}

The rigorous mathematical formalism for the quantitative interpretation
of the DVCS process is based on QCD (Quantum Chromo-Dynamics). The process is illustrated in Fig.~\ref{fig:dvcsbh}-left.
The theory states that the process 
can be factorized between the elementary, precisely calculable, 
photon-quark Compton scattering and some universal structure functions,
called Generalized Parton Distributions (GPDs), which encode the correlations in spatial and momentum distributions 
of the quarks in the proton.
We refer the reader to Refs.~\cite{Mueller:1998fv,Rady96a,Ji97a,Ji97b}
for the original articles on GPDs and to Refs.~\cite{Goeke:2001tz,Diehl:2003ny,
Belitsky:2005qn,Boffi:2007yc,Guidal:2013rya} for recent reviews of the field.
The factorization for the DVCS process has been shown to hold for sufficiently 
large $Q^2$, the squared momentum
transfer between the final and initial leptons, and sufficiently small
$-t \ll Q^2$, the squared momentum transfer between the final 
and initial protons. 

The GPDs are functions of three variables: $x$, $\xi$, and $t$. In a fast moving proton consisting of near-collinear partons, $x+\xi$ ($x-\xi$) 
represent the longitudinal momentum fractions of 
initial (final) quark w.r.t. the average nucleon momentum. The momentum transfer $t$ is the conjugate variable
of the localization of the quark in the transverse position 
plane, perpendicular to the proton momentum direction. 
An interpretation of GPDs thus emerges as distributions describing a quark being 
taken out of the proton with momentum fraction $x+\xi$
and being reinserted in the proton with momentum 
fraction $x-\xi$ at a given 
transverse distance.

Accessing GPDs from DVCS observables is a very challenging task. 
The first challenge results from the fact that in the QCD leading-twist framework in which this work is placed, 
there are four quark helicity-conserving GPDs, denoted by 
$H$, $E$, $\tilde H$, and $\tilde E$, entering the DVCS process.   

A second challenge is that the GPDs actually enter the DVCS amplitude in a form
where they are integrated over $x$. The observables 
thus depend on quantities which are functions of only the two kinematic variables 
$\xi$ and $t$ (neglecting $Q^2$ QCD-evolution effects, given the small $Q^2$ ranges dealt with in this work). 
These observables are called Compton Form Factors (CFFs) and are given for the GPD $H$ by: 
\begin{eqnarray}
{\cal H}_{Re}(\xi , t) &\equiv& {\cal P} \int_0^1 d x  \left\{ \frac{1}{x - \xi} + \frac{1}{x + \xi} \right\} H_+(x, \xi, t), \nonumber \\
{\cal H}_{Im}(\xi , t) &\equiv& H_+(\xi , \xi, t),  
\label{eq:CFFH}
\end{eqnarray}
where $\cal P$ denotes a principal value convolution integral.
The so-called singlet GPD combination $H_+$ is defined as:
\begin{eqnarray}
H^q_+(x, \xi, t) \equiv H^q(x, \xi, t) - H^q(-x, \xi, t). 
\end{eqnarray}

The last complication arises from the fact that there is
another significant mechanism contributing to the $e p \gamma$ final state.
It is the Bethe-Heitler (BH) process, where the final state photon
is radiated by the incoming or scattered electron. 
The process is illustrated in Fig.~\ref{fig:dvcsbh}-right. The BH and 
DVCS mechanisms interfere at the amplitude level. However, the BH amplitude is precisely
calculable within Quantum Electrodynamics  (QED). The only non-QED inputs in the calculation are 
the proton form factors (FFs) $F_1(t)$ and $F_2(t)$ and these are well-known at the small momentum 
transfers $t$ considered in this work.

In Refs.~\cite{Guidal:2008ie,Guidal:2009aa,Guidal:2010ig,Guidal:2010de,Boer:2014kya},
we proposed and applied a method to extract the CFFs from $e p \gamma$ observables. 
It consists in taking the eight CFFs as free parameters
and, knowing the well-established BH and DVCS leading-twist amplitudes, to fit simultaneously
with a least-square method, several $ep\to ep\gamma$ observables, 
at a fixed ($\xi$, $t$) kinematics. In general, an experimental observable receives 
contributions from several CFFs and
there are important correlations between these. The extraction of eight CFFs 
from only a few observables, with finite experimental uncertainties, 
is thus in general an underconstrained problem.
However, some observables are dominated and mostly sensitive to one
or two CFFs. 
For instance, it is well-established 
that the beam-spin observables are 
dominated by the ${\cal H}_{Im}$ CFF. 
Then, if the range of variation of the CFFs is limited, the CFFs dominantly contributing to the observables 
can come out of the fit procedure with finite error bars. These error bars, 
defined by $\Delta\chi^2=+1$ around the minimum $\chi^2$ point, are then in general
due to the correlations between the CFFs. Rather than the
error of the experimental data, they reflect the influence of the other (subdominant) CFFs. Up to the limits 
imposed on the variation of the CFFs,
which should be taken as conservatively as possible, this approach has the merit of being essentially model-independent as 
there is no need to assume and hypothesize any functional shape for the CFFs. This fitting method
was applied, in our earlier works, to derive limits 
and constraints for the ${\cal H}_{Im}$, ${\cal \tilde H}_{Im}$ and ${\cal H}_{Re}$ CFFs at an average $\approx$ 40\% level 
for earlier $ep\to ep\gamma$ data from JLab~\cite{Guidal:2008ie,Guidal:2010ig} 
and HERMES~\cite{Guidal:2009aa,Guidal:2010de}.

Recently, the CLAS and Hall A collaborations of JLab,
using a 5.75 GeV electron beam, have
released new measurements of four observables of the $e p\to e p \gamma$ reaction: 
unpolarized cross sections,  differences of beam-polarized
cross sections (Hall A~\cite{Defurne:2015kxq}
and CLAS~\cite{Jo:2015ema}), 
longitudinally polarized target single spin asymmetries and
double spin asymmetries with both beam and longitudinal target 
polarizations (CLAS~\cite{Pisano:2015iqa,Seder:2014cdc}).
These new data make up the largest set of $ep\to ep\gamma$ observables 
available to date in terms of kinematical coverage and binning.
We have analyzed with the fitting approach outlined above simultaneously all these new data. 

\begin{figure}
\begin{center} 
\vskip -5cm
\includegraphics[width=10 cm]{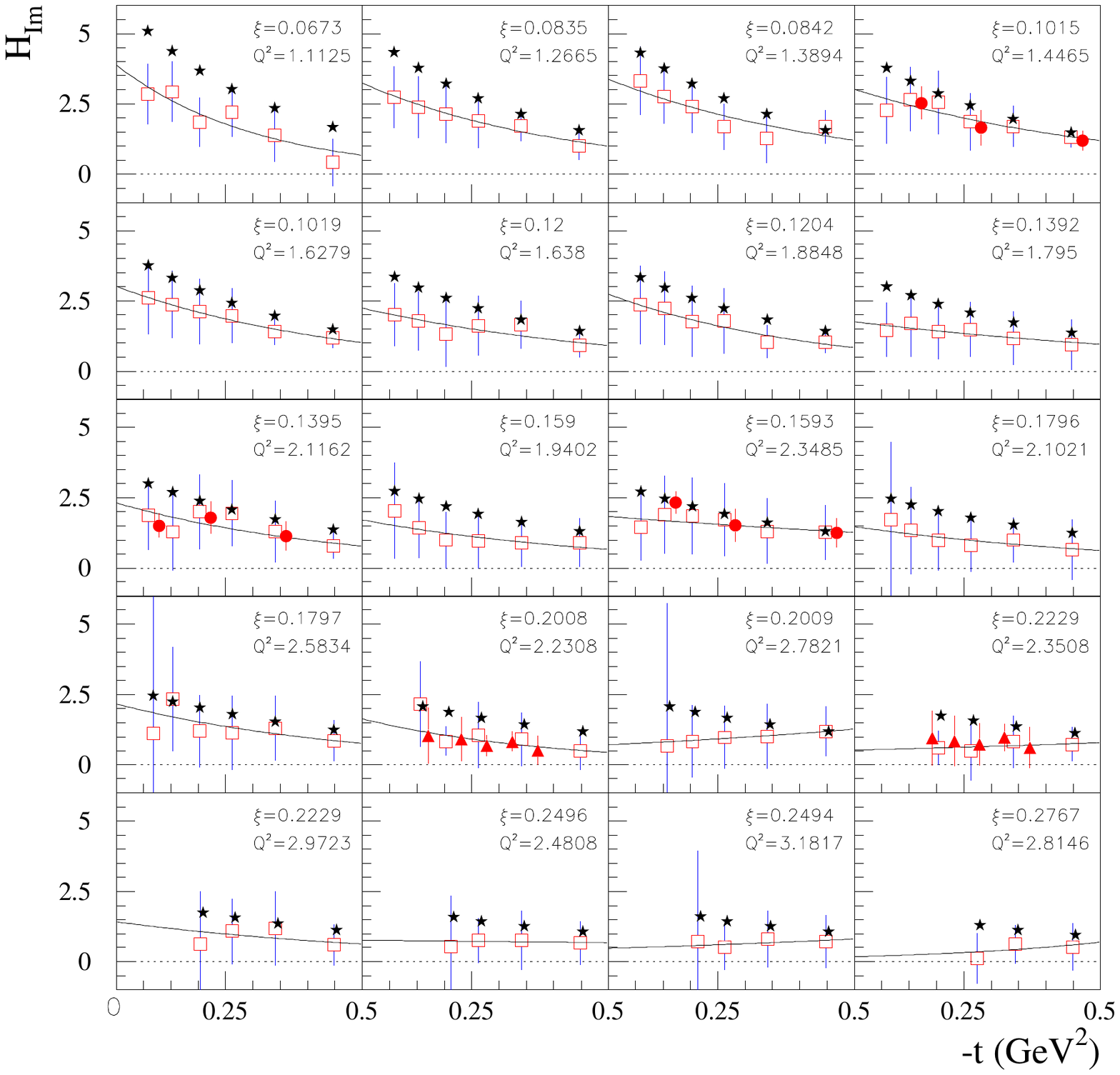}
\caption{$t$-dependence of the CFF ${\cal H}_{Im}$ for 20 CLAS ($x_B$, $Q^2$) bins.
Open squares: results of the CLAS $\sigma$ and $\Delta\sigma$ fit, with the 8 CFFs 
as free parameters. Solid circles: results of the fit to CLAS $\sigma$ and $\Delta\sigma$ data,
as well as longitudinally polarized target and double beam-target polarized asymmetries, with the 8 CFFs 
as free parameters. Solid triangles: results of the Hall A $\sigma$ and $\Delta\sigma$
fit with the 8 CFFs as free parameters. Stars: VGG reference CFFs. 
The solid curve shows an exponential fit of the open squares.}
\label{fig:fithallb_all}
\end{center}
\end{figure}

We focus here on the ${\cal H}_{Im}$ CFF which is the dominant contributor to
the aforementioned JLab observables and which thus comes the most straightforwardly 
and systematically out of the fit with the 
smallest error bars. We show in Fig.~\ref{fig:fithallb_all} the results 
that we obtain at each ($\xi$, $Q^2$, $t$) bin,
from the fit of the JLab CLAS data with 8 CFFs as free parameters. 
Like in our previous works, we have defined the range of variation of the CFFs  
as $\pm 5$ times the CFFs given by 
the VGG model~\cite{Goeke:2001tz,Vanderhaeghen:1998uc,Vanderhaeghen:1999xj,Guidal:2004nd}.
Our fitting procedure has been checked at length and validated by numerous 
Monte-Carlo studies: we generated random 8-CFFs sets, calculated from them observables in a realistic way, 
i.e. smearing these pseudo-data so as to mimick the experimental resolution
of the real data, fitted them by our least-square method with a series of random starting values 
for the CFFs in order to be biased by particular initial conditions, and finally compared
the results to the originally generated CFF values. 
The intensive technical Monte-Carlo studies 
will be detailed in a more detailed methodological article to come.

In Fig.~\ref{fig:fithallb_all}, the results of the 
fit of the CLAS $\sigma$ and $\Delta\sigma$ data are shown by the empty squares. 
For a few ($\xi$, $Q^2$, $t$) bins, longitudinally
polarized target and double beam-target polarized asymmetries from the CLAS experiment are also
available at approximately the same kinematics as the data for $\sigma$ and $\Delta\sigma$. 
We show the values of ${\cal H}_{Im}$ obtained from the simultaneous fit of these 4 observables
with the solid circles. The solid circles have smaller error bars than the 
empty squares as expected, since additional observables in the fit obviously bring 
new constraints. 
We also added in Fig.~\ref{fig:fithallb_all} the result of the fits of $\sigma$ and $\Delta\sigma$ 
from Hall A, where there is overlap with the CLAS data. 
There is in general a good agreement
between the ${\cal H}_{Im}$ values extracted from both experiments. 
For reference, we show in Fig.~\ref{fig:fithallb_all} 
the predictions of the VGG model. The comparison shows that the version of the VGG model
that has been taken for the reference CFF (corresponding with $b_v = b_s = 1$) tends to overestimate the data 
at small values of $t$ by around 30 \%.

We observe the general trend that ${\cal H}_{Im}$ decreases with $t$ and that these $t$-slopes
tend to become steeper as $\xi$ decreases. We have quantified this
and extracted a general $(\xi,t)$-dependence of the CFFs, by fitting 
the $t$-dependence with an exponential function as given by:
\begin{eqnarray}
{\cal H}_{Im}(\xi, t) = A(\xi) e^{B(\xi) t}.   
\label{eq:him}
\end{eqnarray}  
The solid lines in Fig.~\ref{fig:fithallb_all} show the results of these exponential fits of the empty squares.

\begin{figure}[h] 
\begin{center} 
\includegraphics[width=5.5cm]{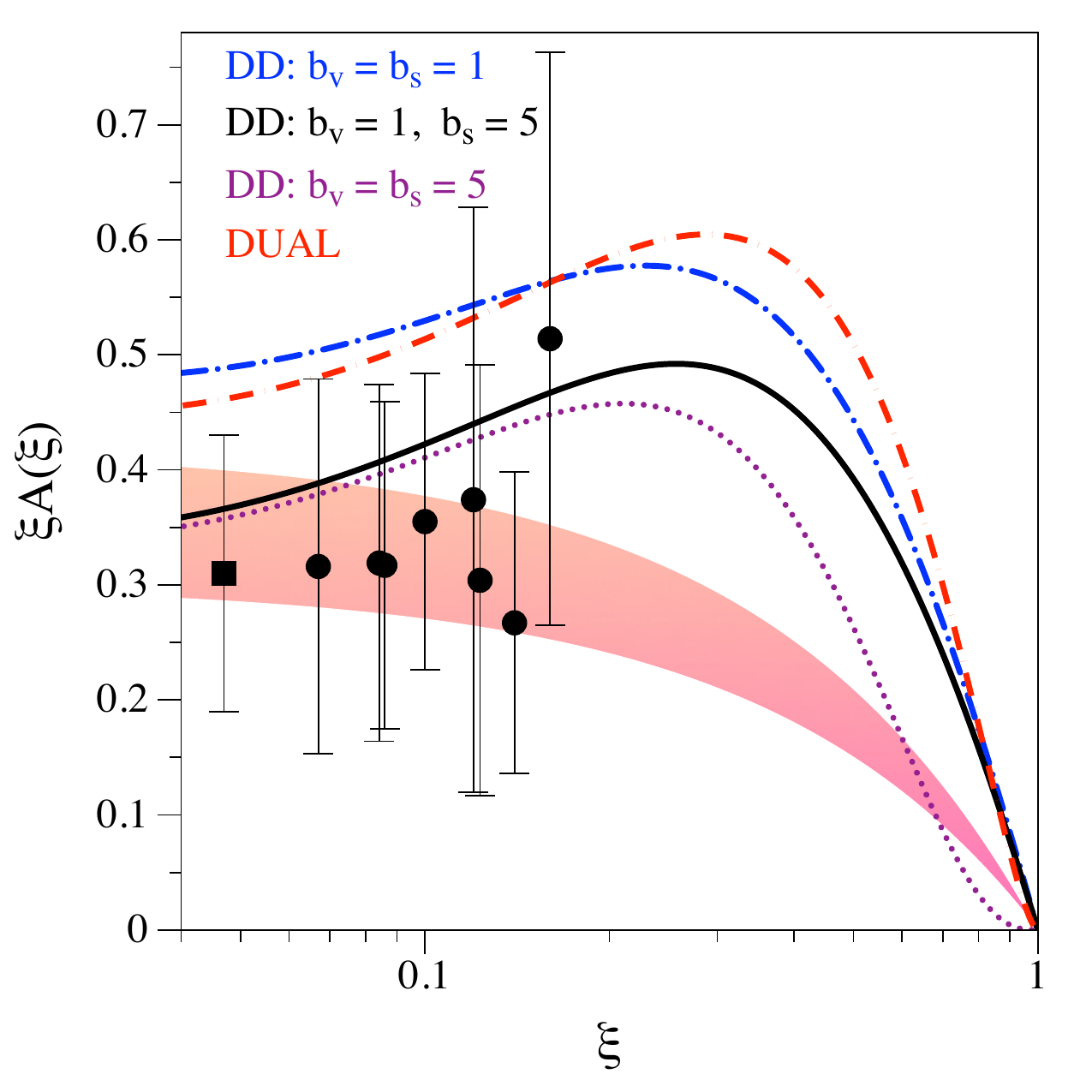}
\includegraphics[width=5.5cm]{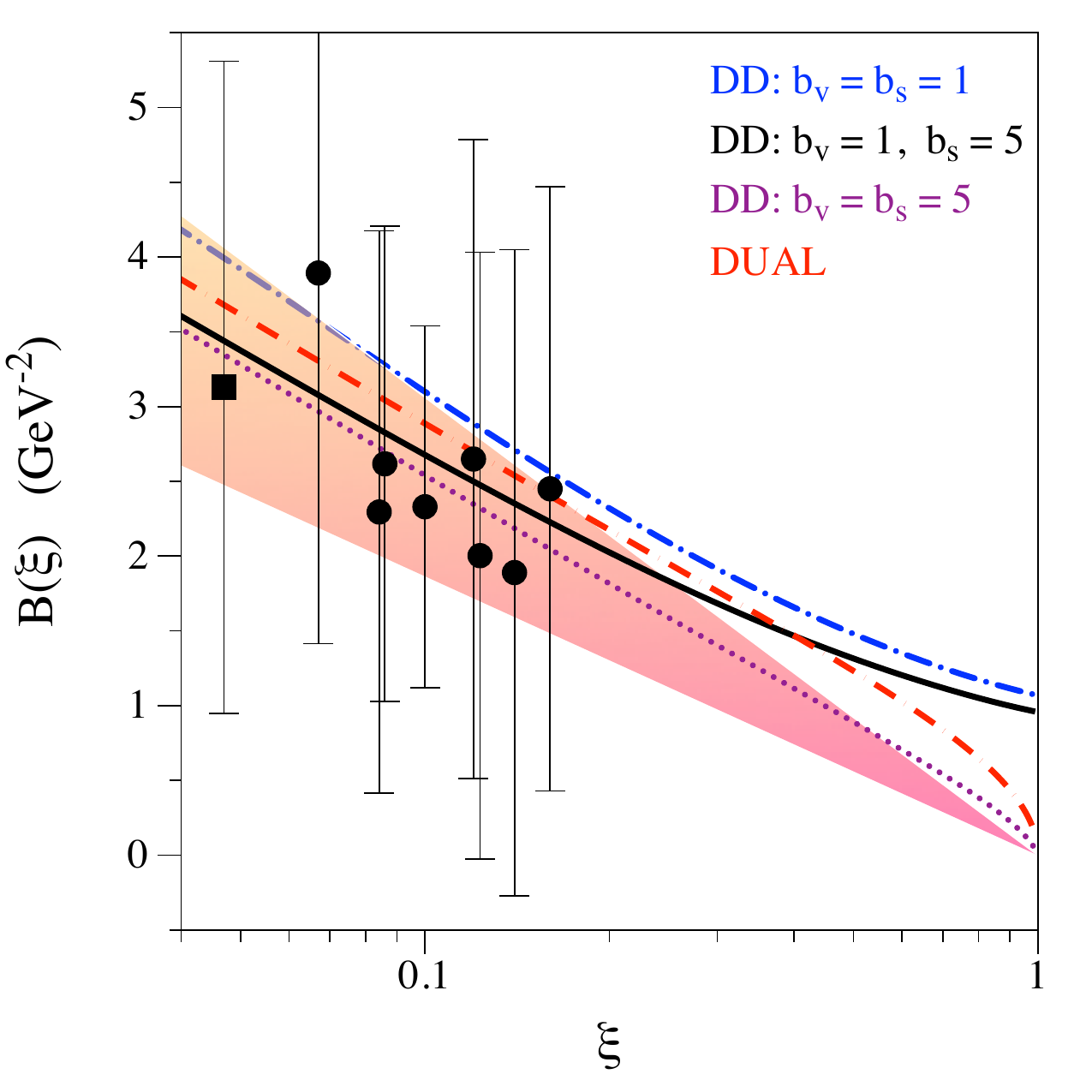}
\caption{Upper: Amplitude $A$ of ${\cal H}_{Im}$, multiplied by $\xi$, 
as a function of $\xi$. Lower: $t$-slope $B$ of ${\cal H}_{Im}$ as a function of $\xi$. 
Data points: 8 CFFs fit of 
data 
corresponding with 4 observables
from JLab/CLAS (circles), and from HERMES (square) 
the latter as extracted in Refs.~\cite{Guidal:2009aa,Guidal:2013rya}. 
For the JLab points, we kept only the 8 lowest $\xi$ bins of Fig.~\ref{fig:fithallb_all}
which provide the lowest uncertainty results.  
The one-parameter fits to these data points according to Eqs.~(\ref{eq:fita}, \ref{eq:fitb}) are shown by the bands,  corresponding with a $1 \sigma$ variation of $a_A$ and $a_B$, given by Eq.~(\ref{eq:fitAB}).    
The theory curves  correspond with the dual model and the double distribution (DD) model for three choices of the  valence (sea) profile parameters $b_v$ ($b_s$), as indicated.    
}
\label{fig:fita}
\end{center}
\end{figure}


In Fig.~\ref{fig:fita}, we plot the dependence of both the amplitude $A$ and the exponential $t$-slope $B$ on $\xi$. 
In spite of the large size of the errors, which are not statistical
we recall, one can observe that, systematically, both $A$ 
and $B$ tend to increase as $\xi$ decreases.
Physically, at small $\xi$, one expects $A$ to 
to rise steeply as $1 / \xi$ due to the sea-quark contribution. Furthermore, $A$  
is expected to vanish in the limit $\xi \to 1$, when one valence quark takes all longitudinal momentum. Therefore, we fit $A$ by the simple one-parameter form which embodies both features: 
\begin{eqnarray}
A(\xi) &=& a_A (1 - \xi)/ \xi, 
\label{eq:fita}
\end{eqnarray} 
and will extract the parameter $a_A$ from a fit to the data. 
For the slope $B$, we expect it to sharply decrease from a Regge type behavior when $\xi \to 0$ to 
a flat $t$-dependence in the limit $\xi \to 1$, reflecting the pointlike coupling to a valence quark carrying all longitudinal momentum. 
To encompass both limits, we fit the slope $B$ by the following one-parameter ansatz in $\xi$:  
\begin{eqnarray}
B(\xi) = a_B \, \mathrm{ln}(1/\xi). 
\label{eq:fitb}
\end{eqnarray} 
The rise of $B$ at small $\xi$ corresponds to the increase of the
transverse size of the proton as smaller longitudinal momentum fractions
are probed. 
A fit to the data with the functional forms of Eqs.~(\ref{eq:fita}, \ref{eq:fitb}) yields the values:
\begin{eqnarray}
a_A =  0.36  \pm 0.06 , \quad 
a_B &=& 1.07 \pm 0.26 \; \mathrm {GeV}^{-2}.
\label{eq:fitAB}
\end{eqnarray}
The resulting fits are shown by the bands in Fig.~\ref{fig:fita}.


We can confront the experimentally extracted values of $A$ and $B$ with the expectations from GPD models, as shown in Fig.~\ref{fig:fita}.  
We compare two GPD models: the dual model~\cite{Polyakov:2008xm} and the VGG double distribution (DD) model~\cite{Vanderhaeghen:1998uc,Vanderhaeghen:1999xj,Goeke:2001tz,Guidal:2004nd}.
For the latter, we use three choices of the valence (sea) profile parameters $b_v$ ($b_s$) respectively. 
For large values of these profile parameters 
($b \to \infty$), the GPD $H(x,\xi,t)$ tends to the GPD $H(x,0,t)$, where the 
effect of the skewness ($\xi$-dependence) disappears. 
For the dual model, we have used the lowest forward-like function. 
For both models, we use the same empirical forward parton distributions as input and use in both cases a 
Regge parameterization for the $t$-dependence with slope parameter $1.05$~GeV$^{-2}$, see 
Ref.~\cite{Guidal:2013rya} for details.
 
Comparing the extracted data for $A$ with theory, we notice from Fig.~\ref{fig:fita} that in the region $0.05 \lesssim \xi \lesssim 0.2$ the data tend to lie systematically below the result of the dual model (with lowest forward-like function), as well as the DD models where sea quarks have strong skewness ($b_s = 1$). The DD models with small skewness effects of sea-quarks ($b_s = 5$) are in good agreement with the data. To distinguish for the valence quarks between the cases of strong skewness ($b_v = 1$) and weak skewness ($b_v = 5$) will require data in the region 
$\xi \gtrsim 0.3$. We also notice from Fig.~\ref{fig:fita} 
that the GPD models predict a maximum for $\xi A(\xi)$ around $\xi \approx 0.3$, due to the $x$-dependence of the underlying valence quark distributions. 

In the lower panel of Fig.~\ref{fig:fita}, 
we show the exponential $t$-slope $B(\xi$), as a  function of $\xi$. 
We notice that all GPD models, which are based on a Regge parameterization for its $t$-dependence,
are in good agreement with the available data for $B$. 
Both the data as well as the models follow a $\ln (1/\xi)$ behavior, thus leading to an increase of the slope as $\xi$ decreases. Only for $\xi \gtrsim 0.5$, some qualitative differences between the models appear. 

We now seek to relate the increasing $t$-slope $B(x)$ when $x$ decreases with the 
variation of the spatial size of the proton when probing partons with different longitudinal momentum fraction $x$. 
For this purpose, we relate it to the (helicity averaged) transverse charge distribution in the proton, denoted by $\rho$, 
which is obtained through a 2-dimensional 
Fourier transform of the FF $F_1$ as~\cite{Burkardt:2000za}:
\begin{equation}
\rho({\bf b_\perp})=\int
\frac{d^2 {\bf \Delta_\perp}}{(2\pi)^2}e^{- i {\bf b_\perp \cdot \Delta_\perp}} 
F_1(-{\bf \Delta}^2_\perp),
\label{eq:fourier1}
\end{equation}
where $\bf b_\perp$ denotes the quark position in the plane transverse to the longitudinal momentum of a fast moving proton, and ${\bf \Delta_\perp}$ 
denotes the transverse components of the momentum transferred to the proton.
The squared radius of the unpolarized 2-dimensional transverse charge distribution in the proton is then defined as:
\begin{equation}
\langle b^2_\perp \rangle  = \int d^2 {\bf b_\perp} {\bf b}^2_\perp \rho({\bf b_\perp}). 
\label{eq:cr1}
\end{equation}
The quantity  $\langle b^2_\perp \rangle$ is related to the conventionally defined squared radius $\langle r_1^2 \rangle$ of the proton FF $F_1$ 
as $\langle b^2_\perp \rangle  = 2/3 \langle r_1^2 \rangle$.  
The experimental value of $\langle r_1^2 \rangle $ based on elastic electron-proton scattering data 
yields~\cite{Bernauer:2013tpr}: $\langle r_1^2 \rangle = 0.65 \pm 0.01~\mathrm{fm}^2$, 
resulting in the empirical value for the proton's transverse squared radius:
\begin{equation}
\langle b^2_\perp \rangle = 0.43 \pm 0.01~\mathrm{fm}^2 = 11.05 \pm 0.26~\mathrm{GeV}^{-2}.   
\label{eq:cr4}
\end{equation}

Similarly to the FFs, the $t$ variable in the GPDs is the conjugate variable
of the impact parameter. For $\xi=0$ (for which $t=-\Delta_\perp^2$), 
one therefore has an impact parameter version of GPDs through a Fourier integral in $\Delta_\perp$, which for a parton of flavor $q$ reads as~:
\begin{equation}
\rho^q(x, {\bf b_\perp})=\int
\frac{d^2 {\bf \Delta_\perp}}{(2\pi)^2}e^{- i {\bf b_\perp \cdot \Delta_\perp}}H^q_-(x,0,-{\bf \Delta_\perp}^2).
\label{eq:fourier}
\end{equation}
Here $H^q_-(x, 0, t)$ is the so-called non-singlet or valence GPD combination, 
defined as:
\begin{eqnarray}
H^q_-(x,0, t) \equiv H^q(x,0,t) + H^q(-x,0,t),   
\end{eqnarray}
with $0 \leq x \leq 1$.  
At $\xi$=0, the function $\rho^q(x, {\bf b_\perp})$ can then be interpreted as the number density of quarks of flavor $q$ with {\it longitudinal} momentum 
fraction $x$ at a given {\it transverse} distance ${\bf b_\perp}$ 
in the proton~\cite{Burkardt:2000za}. 
Generalizing Eq.~(\ref{eq:cr1}), one can 
define the $x$-dependent squared radius of this quark density in the transverse plane as:
\begin{equation}
\langle b^2_\perp \rangle^q (x) = \frac{ \int d^2 {\bf b_\perp} {\bf b}^2_\perp \rho^q(x, {\bf b_\perp})}{\int d^2 {\bf b_\perp}  \rho^q(x, {\bf b_\perp})},
\label{eq:cr5}
\end{equation}
which can be expressed through the GPD $H_-$ as:
\begin{equation}
\langle b^2_\perp \rangle^q (x)= - 4 \frac{\partial}{\partial {\bf \Delta}^2_\perp} \ln H^q_-(x,0,-{\bf \Delta_\perp}^2) \biggr| _{{\bf \Delta_\perp} = 0}.
\label{eq:crgpd}
\end{equation}
Assuming the $t$-dependence of the valence GPD $H^q_-(x,0,t)$ to be exponential of the form:
\begin{eqnarray}
H^q_-(x,0, t) = q_v(x) e^{B_0(x) t},   
\label{eq:hzeroxi}
\end{eqnarray}
with $q_v(x)$ the corresponding valence quark distribution, 
Eq.~(\ref{eq:crgpd}) then yields for each flavor $q$:
\begin{eqnarray}
\langle b^2_\perp \rangle^q (x)= 4 B_0(x). 
\label{eq:bperp1}
\end{eqnarray}
The $x$-independent squared radius is obtained from $\langle b^2_\perp \rangle^q (x)$ through the following 
average over $x$:
\begin{eqnarray}
\langle b^2_\perp \rangle^q 
=  \frac{1}{N_q} \int_0^1 dx \,q_v(x) \, \langle b^2_\perp \rangle^q (x), 
\label{eq:bperp2}
\end{eqnarray}
with the integrated number of valence quarks $N_u = 2$ and $N_d = 1$.
For the proton, the Dirac squared radius  $\langle b^2_\perp \rangle $ 
is then obtained as the charge weighted sum over the valence quarks:
$\langle b^2_\perp \rangle =  2 e_u \langle b^2_\perp \rangle^u + e_d \langle b^2_\perp \rangle^d$, 
with quark electric charges $e_u = +2/3$ and $e_d = -1/3$.
A Regge ansatz for the $t$-slope of $H_-^q(x,0,t)$ yields: 
\begin{eqnarray}
B_0(x) = a_{B_0} \ln(1/x), 
\label{eq:B0}
\end{eqnarray}
with $a_{B_0}$ the Regge slope. When evaluating the corresponding integral of Eq.~(\ref{eq:bperp2}), 
using the  empirical constraint of Eq.~(\ref{eq:cr4}) for 
$\langle b^2_\perp \rangle$, we obtain the estimate:
\begin{eqnarray}
a_{B_0} = \left(1.05 \pm 0.02 \right) \mathrm{GeV}^{-2} . 
\label{eq:aB0}
\end{eqnarray}


\begin{figure}
\begin{center} 
\includegraphics[width=5.6cm]{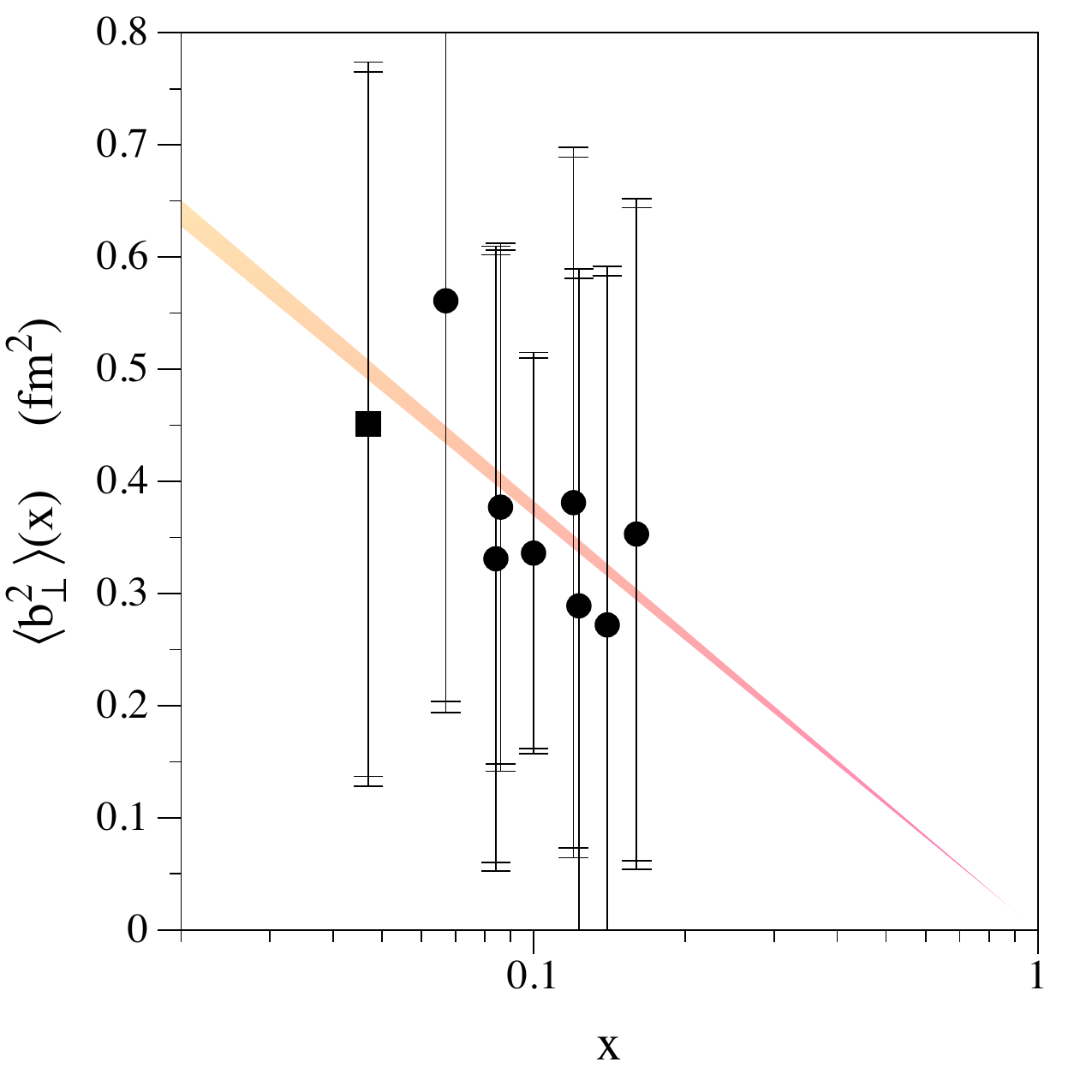}
\caption{$x$-dependence of  $\langle b^2_\perp \rangle$ 
for quarks in the proton. 
The band shows the empirical result using the logarithmic ansatz for $B_0(x)$ 
of Eqs.~(\ref{eq:B0}, \ref{eq:aB0}). The data points correspond with the results obtained in this work for $B(x)$, as displayed in Fig.~\ref{fig:fita}. 
They have been multiplied by the correction factor 
$B_0 / B \simeq 0.925 \pm 0.025$ in the $x$-range of the data. The resulting (small) model uncertainty is shown by the outer error bars.  
}
\label{fig:bperp}
\end{center}
\end{figure}

\begin{figure}
\begin{center} 
\includegraphics[width=7.5cm]{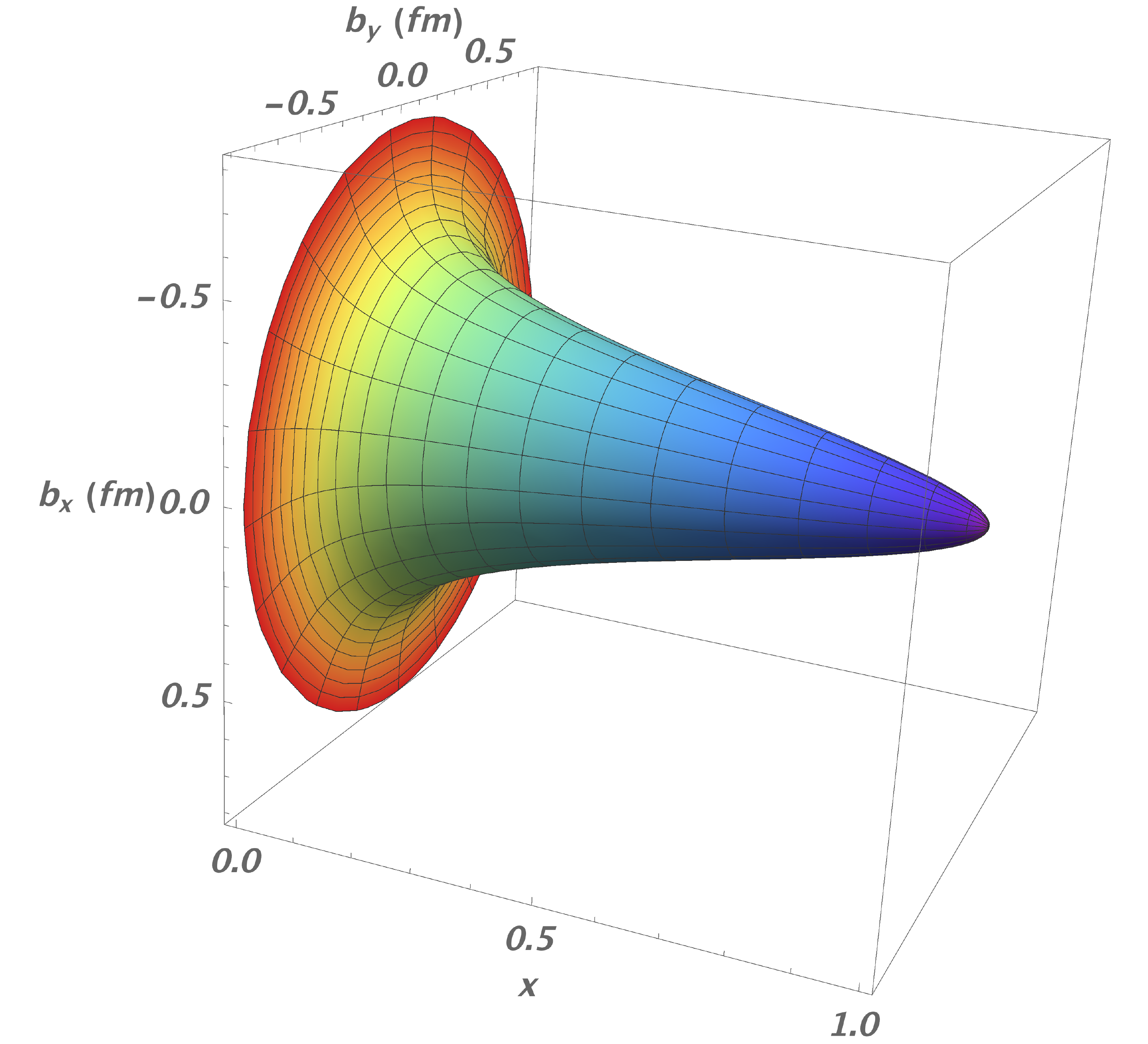}
\caption{$x$-dependence of the proton's transverse extent obtained from the data of Fig.~\ref{fig:bperp}.  
}
\label{fig:illus}
\end{center}
\end{figure}

To quantitatively compare this with the $t$-slope of ${\cal H}_{Im}$ defined through Eq.~(\ref{eq:him}), 
we need to be aware of a difference. 
The experimentally measured $t$-slope $B(x)$ is for the singlet GPD combination $H_+(x,x,t)$. 
On the other hand, the $t$-slope $B_0(x)$ of Eq.~(\ref{eq:B0}, \ref{eq:aB0}) 
is for the valence GPD in the limit $\xi = 0$, i.e. for the function 
$H^q_-(x,0, t)$ for a quark of flavor $q$. In our analysis, we assume that the function $B_0(x)$ is the same for $u$ and $d$ quarks, in agreement with the observed universality of the Regge slopes for meson trajectories. 
To get some quantitative idea how large the difference between the (flavor independent) slopes 
$B_0$ and $B$ is, we have studied the $x$-dependence of the ratio $B_0(x) / B(x)$ within both the dual and DD GPD models.
For the $x$ range of the available data,   $0.05 \lesssim x \lesssim 0.2$, we notice that 
the GPD models with $b_s = 5$, which were found to be compatible with both the data for $A$ and $B$, 
yield:
$0.90 < B_0/B < 0.95$. 
As a result, we can convert the data for $B(x)$ to data for  
$\langle b^2_\perp \rangle(x)$  using Eq.~(\ref{eq:bperp1}), as shown in 
Fig.~\ref{fig:bperp}. They are compared with the result using the logarithmic ansatz for $B_0(x)$ 
of Eq.~(\ref{eq:B0}), with parameter $a_B$ 
determined from the proton Dirac radius. 
One sees that within errors both determinations are perfectly compatible.  We have here extracted the $x$-dependence of the squared radius of the quark distributions in the transverse plane, demonstrating an increase of this radius with decreasing value of the longitudinal quark momentum fraction $x$. 
Fig.~\ref{fig:illus} shows a three-dimensional view of the numerical function that we obtained by the 
fit of the data of Fig.~\ref{fig:bperp}. 

In summary, we have analyzed in a GPD QCD leading-twist and leading-order framework the latest $e p\to e p \gamma$ unpolarized cross sections, difference of beam-polarized cross sections, longitudinally polarized target single spin,
and beam-longitudinally polarized target double spin asymmetries recently measured
at JLab. We have extracted constraints on the 
${\cal H}_{Im}$ CFF over a large range in $\xi$.  From the amplitude and the $t$-slope of ${\cal H}_{Im}$,
we have been able to derive a functional mapping of the density and transverse size of the proton charge  
as a function of the quark's longitudinal momentum.

\section*{Acknowledgments}

We are very thankful to S. Niccolai and D. Mueller for very useful
discussions. R.D. and M.G. benefitted from the ANR-12-MONU-0008-01 ``PARTONS" contract support. 
The work of M.V. was supported in part by the Deutsche Forschungsgemeinschaft DFG through the Collaborative Research Center SFB 1044. 




\end{document}